\documentclass[%
 letterpaper,
 reprint,
 superscriptaddress,
nofootinbib,
 amsmath,amssymb,
 aps,
 prd,
]{revtex4-2}

\usepackage{mathptmx}
\usepackage{graphicx}\graphicspath{{figures/latest/}{figures/}}
\usepackage{bm}
\usepackage[table]{xcolor}
\definecolor{darkred}{rgb}{0.5,0,0}
\definecolor{darkblue}{rgb}{0,0,0.5}
\definecolor{firebrick}{rgb}{0.75,0.125,0.125}
\definecolor{darkgreen}{rgb}{0,0.5,0}
\usepackage[colorlinks=true,linkcolor=darkgreen,urlcolor=darkblue,citecolor=darkgreen]{hyperref}
\usepackage{upgreek}
\usepackage[capitalise]{cleveref}
\usepackage{xfrac}
\usepackage{multirow}

\def\dd{\mathrm{d}}

\begin{document}

\preprint{APS/123-QED}

\title{Neutron Production in Simulations of Extensive Air Showers}

\def\iap{\affiliation{Institute for Astroparticle Physics, Karlsruhe Institute of Technology \includegraphics[height=1.55ex]{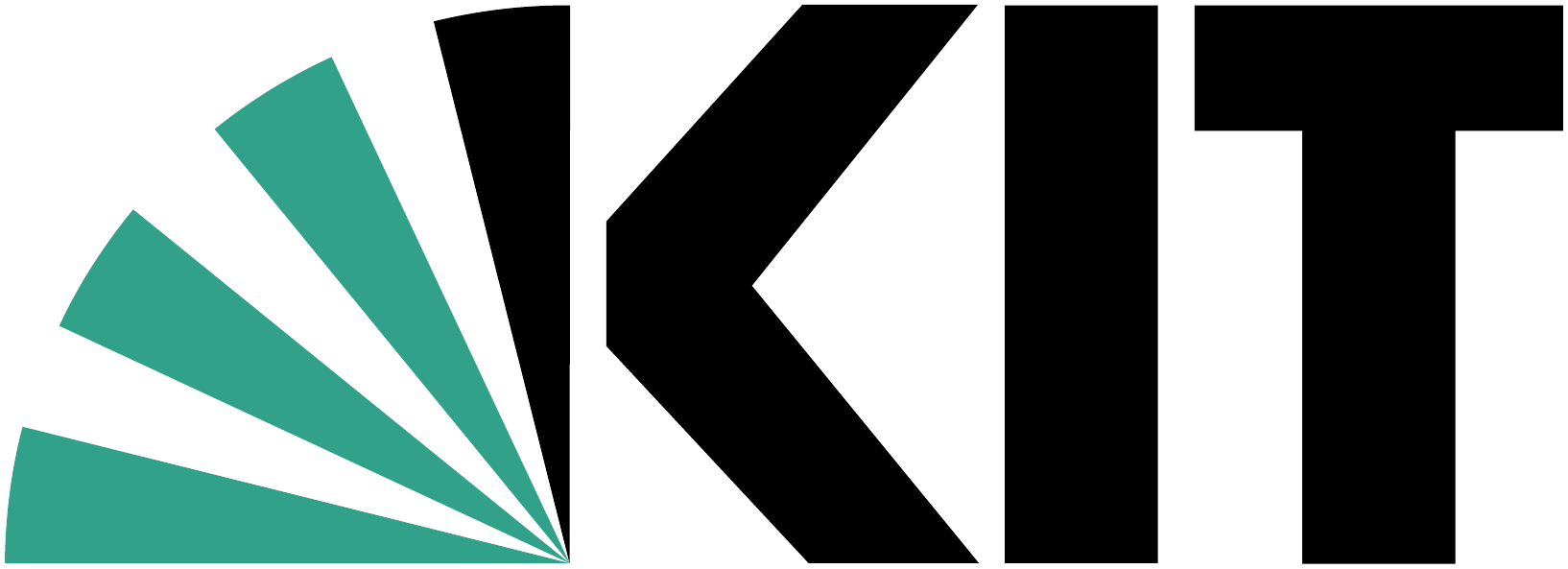}, Germany}}
\def\ietp{\affiliation{Institute of Experimental Particle Physics, Karlsruhe Institute of Technology \includegraphics[height=1.55ex]{kit_logo-no_text}, Germany}}
\def\ijc{\affiliation{CNRS/IN2P3, IJCLab \raisebox{-0.2ex}{\includegraphics[height=2.3ex]{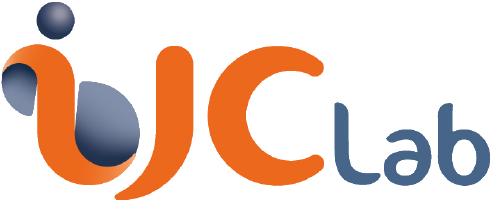}}, Universit\'e Paris-Saclay, Orsay, France}}

\iap
\ietp
\ijc

\author{Martin Schimassek}\email{martin.schimassek@ijclab.in2p3.fr}
\iap
\ijc

\author{Ralph Engel}\email{ralph.engel@kit.edu}
\iap
\ietp

\author{Alfredo Ferrari}\email{alfredo.ferrari@kit.edu}
\iap

\author{Markus Roth} 
\iap

\author{David Schmidt}
\ietp

\author{Darko Veberi\v{c}}
\iap

\date{\today}

\begin{abstract}
Although the electromagnetic and muonic components of extensive air showers have been studied in great detail, no comprehensive simulation study of the neutron component is available.
This is related to the complexity of neutron transport processes that is typically not treated in standard simulation tools.
In this work we use the Monte Carlo simulation package \textsc{Fluka} to study the production and the transport 
of neutrons in extensive air showers over the full range of neutron energies, extending down to thermal neutrons. 
The importance of different neutron production mechanisms and their impact on predicted neutron distributions in energy, lateral distance, atmospheric depth, and arrival time are discussed.
In addition, the dependencies of the predictions on the properties of the primary particle are studied.
The results are compared to the equivalent distributions of muons, which serve as reference.
\end{abstract}

\maketitle


\section{Introduction}
\label{sec:intro}

In extensive air showers, neutrons are the only neutral and stable\footnote{Stable on time scales of relevance for air-shower physics.} hadronic particles produced.
Since they do not lose energy through ionization, their propagation and energy losses in the atmosphere are determined only by hadronic interactions and quasielastic scattering processes.
In addition, there is a unique production process of slow neutrons, the spallation of air nuclei with the subsequent emission of evaporation neutrons.
As a result, the neutron cloud of an air shower has properties that are fundamentally different from those of the electromagnetic and muonic shower components.
At the same time, the neutron component offers an observable linked to hadronic processes in air showers, similar to the muons.

Studies of the neutron component of showers have been carried out since the early years of air shower measurements, for example, Ref.~\cite{PhysRev.75.1532}. 
It was already clear at that time that there was a considerable time delay between the arrival of the low-energy neutrons at the ground and the arrival of the other shower particles, and this delay was used for their identification. 
In a seminal paper, Linsley discussed the observation of late pulses in a scintillator air-shower array, caused by so-called subluminal particles, and estimated the energy content of the neutron component of air showers~\cite{Linsley:1984bnz}. 
Today we have a number of measurements of slow neutrons produced in air showers, made with dedicated detectors, see for example Refs.~\cite{Shepetov:2019urg,Stenkin:2007zz,Erlykin:2006qw} for a discussion. 
However, the difficulty of computing reliable predictions and the dependence of these predictions on details of the environment in which they are observed complicate the interpretation of neutron data.
Moreover, the detection of neutrons requires specialized types of detectors, with which it is difficult to cover large areas as typically needed for air-shower physics.
Therefore, it is not surprising that modern air-shower arrays are not built to detect neutrons and that typical air-shower simulation programs are not designed to handle low-energy neutrons.

In this article, we revisit the production of neutrons in air showers, motivated by the operation or planned installation of large scintillator arrays which are expected to have some, albeit limited, sensitivity to neutrons. 
We use \textsc{Fluka}~\cite{Ferrari:2005zk,Bohlen:2014buj} to simulate air showers in detail, fully considering the production and propagation of neutrons of all energies in the atmosphere. 
Based on the simulation results, we present general features of the neutron component in air showers that are independent of possible detection techniques. 

The structure of the article is as follows. 
In \cref{sec:fluka} we demonstrate that \textsc{Fluka} is a well-suited tool for carrying out neutron studies by comparing \textsc{Fluka} predictions with dedicated neutron measurements.
The method to simulate the air showers analyzed in this work is introduced in \cref{sec:simmethod}.
As we are interested in understanding the fundamental aspects of neutron production in air showers, here we consider only vertical showers in the energy range of $10^{14}$ to $10^{18}$,eV, and only protons, Fe nuclei, and photons as primary particles.
It is beyond the scope of the present work to perform a comprehensive survey of shower energies or any other parameters, since such studies must be performed in combination with dedicated simulations of specific experimental setups in order to be useful.
A selection of the obtained neutron distributions is shown in \cref{sec:results},
The neutron distributions are used here to derive a number of key observations.
To put the results into a more general perspective, a comparison is made with muon distributions, obtained in the same showers, where applicable.
In addition, a first estimate of the detection efficiency of neutrons of different energies is presented for generic scintillator and water-Cherenkov detectors.
The article concludes with a brief discussion in \cref{sec:discussion}.

\section{The \textsc{Fluka} code}
\label{sec:fluka}

\textsc{Fluka}~\cite{fluka:url} is a fully integrated Monte Carlo simulation package for interaction and transport of particles and nuclei in matter.
\textsc{Fluka} has many applications in particle physics, high-energy experimental physics and engineering, shielding, detector design, cosmic-ray studies, dosimetry, medical physics, radiobiology, and hadron therapy.
Over the last 30 years, there have been a multitude of verifications of the accuracy of the \textsc{Fluka} code in predicting particle production spectra, and in particular neutrons, around high-energy accelerators, and by cosmic rays.

\subsection{Physics processes in \textsc{Fluka}}
\label{subsec:flukaPhysics}

\textsc{Fluka} nuclear reaction models are based on the \textsc{Peanut}~\cite{Bohlen:2014buj,Ref8,Fasso:2001} hadron-nucleus generator for projectile energies up to 20\,TeV in the laboratory system. 
At higher hadron energies, the latest development~\cite{Ref9} of the \textsc{DPMJet-III} code~\cite{Ref10} is used: this code has been extensively benchmarked against LHC data and its predictions are therefore tuned to collider data until up to tens of TeV in the nucleon-nucleon center-of-mass. 
For ion projectiles, \textsc{DPMJet-III} is again used for energies above 5\,GeV/$A$, while at lower energies an extensively modified~\cite{Cospar2002} version of rQMD-2.4~\cite{Sorge:1995dp} is used for nucleus-nucleus interactions. 
Finally, of minimal relevance for this study, \textsc{Fluka} uses the BME model~\cite{Ref12} for ions below 125\,MeV/$A$. 
\textsc{Fluka} can also deal with real and virtual photonuclear interactions, a capability that is critical to the results presented in this paper.
Among virtual photon interactions, electromagnetic dissociation in ion-ion collisions is properly described and has been validated up to LHC center-of-mass energies~\cite{Braun:2014naa}.
Photodissociation of target nuclei is a potentially important process for the production of low-energy (i.e.\ slow) neutrons.

For neutrons below 20\,MeV and down to $10^{-5}$\,eV, two different approaches can be used in \textsc{Fluka}, both based on evaluated nuclear-data files suitably processed for \textsc{Fluka}.
The former makes use of multigroup neutron cross sections obtained by processing the original evaluated data with \textsc{Njoy2016}~\cite{njoy2016:url} into 260 energy groups, out of which 30 cover the thermal energy range.
The latter approach makes use of the same evaluated data processed into continuous pointwise cross sections by means of \textsc{Prepro19}~\cite{prepro19:url} and \textsc{Njoy2016}: this approach is recommended for medium-heavy materials with cross sections with complex resonance structures, or when correlated, fully exclusive low-energy neutron interactions are of interest.
More details on both approaches can be found in the \textsc{Fluka} documentation~\cite{fluka:url}.
For the present paper, the two approaches are equivalent since the cross sections of the main air components do not exhibit extensive resonance regions, therefore the faster multi-group approach has been used.
For the air isotopes, the basic cross section data have been derived from the \textsc{Endf/b-VIIIr0} database~\cite{Endfb8r0}.

An example of the accuracy, which can be achieved in predicting neutron spectra in a well-controlled environment where all details of the incoming beam, the geometry, and materials are well-known, can be found in Ref.~\cite{NTOF}.
In that article, the code predictions are compared with the neutron spectrum measured at the n\_TOF facility at CERN (see Figs.~5 and 6 of Ref.~\cite{NTOF}, and the remark about the actual thickness of the moderation layer).

\subsection{Validation of \textsc{Fluka} for cosmic-ray calculations}
\label{subsec:flukagcr}

In the literature, several validations of \textsc{Fluka} for cosmic ray calculations exist: Comparisons of code predictions with measured muon spectra in the atmosphere can be found in Refs.~\cite{Battistoni:2001fp,Battistoni:2007zza}, and for proton and neutron spectra in Refs.~\cite{Zuccon:2003ns,PELLIC1,COMBIER}.
Recent applications/validations of the code related to the energy range of Galactic cosmic rays can be found in Refs.~\cite{Fermi-LAT:2016tkg,Mazziotta:2020uey}.

\begin{figure*}
\centering
\def\w{0.485}
\includegraphics[width=\w\linewidth]{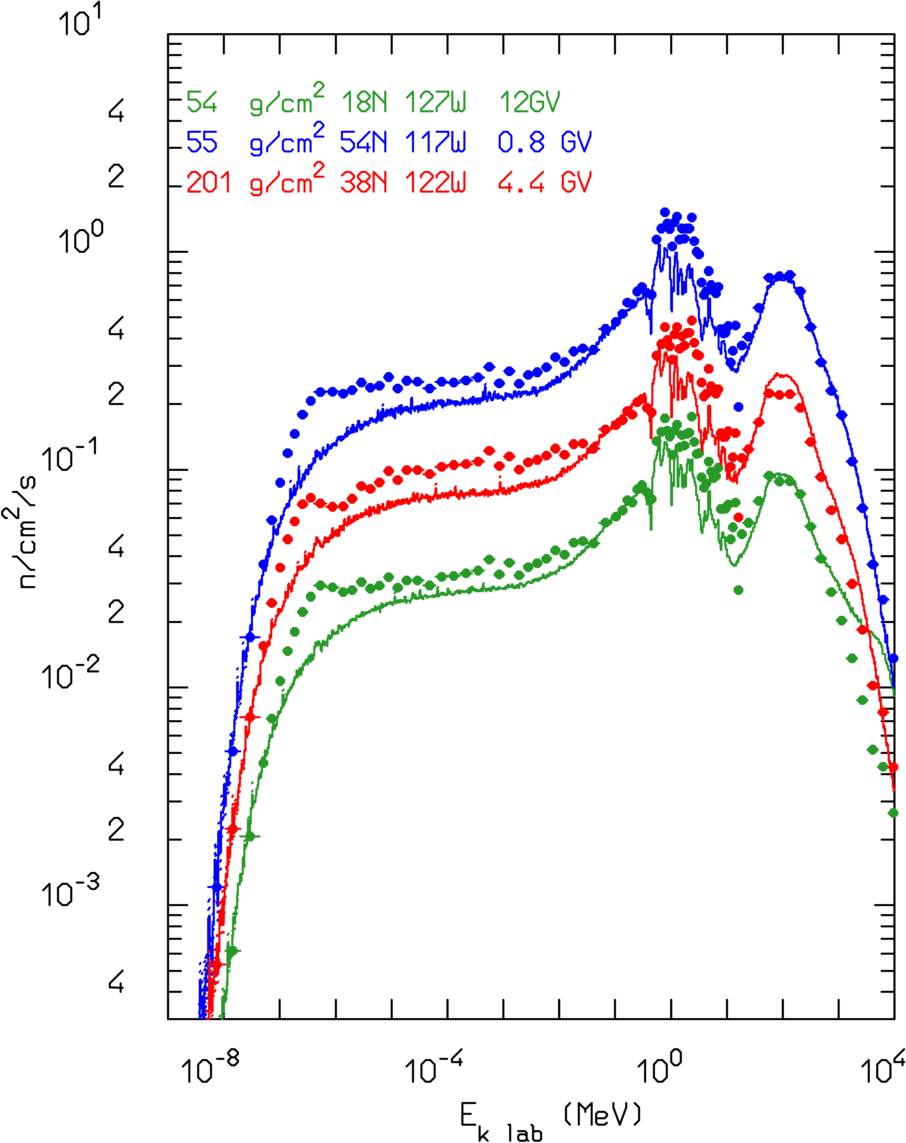}\hfill
\includegraphics[width=\w\linewidth]{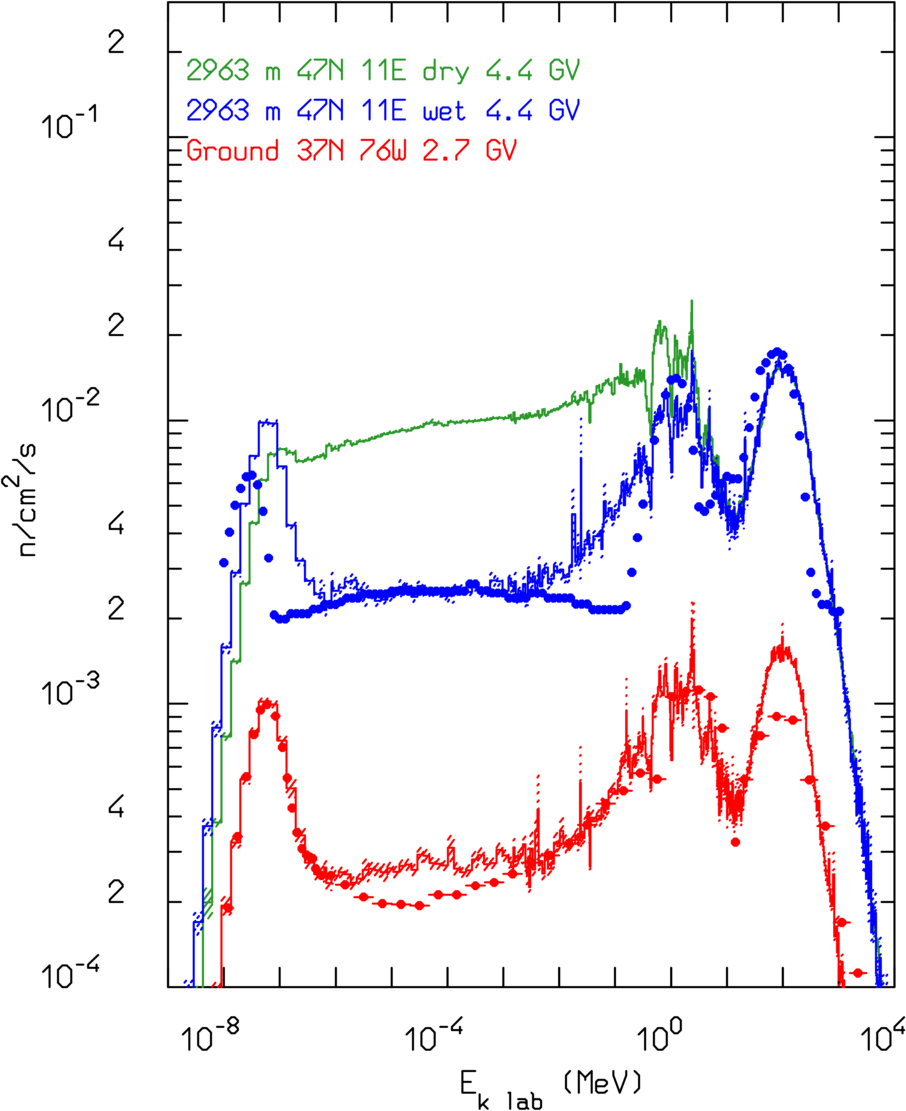}
\caption{\textsc{Fluka} (line) and experimental (symbols) neutron spectra in the atmosphere:
\emph{left:} at three different depths and locations, data from Ref.~\cite{Ref13},
\emph{right:} at top of the Zugspitze mountain (altitude 2963\,m)~\cite{Ref14} (green for a completely dry layout, blue under the assumptions described in the text), and at ground level~\cite{Ref13} (red).
The location latitudes (N\,=\,North), and longitudes (E\,=\,East, W\,=\,West) are indicated as well as the vertical rigidity cutoff for the measurements taken at fixed locations. 
The dashed areas represent the statistical errors of the calculation.}
\label{fig:1}
\end{figure*}

To illustrate the capabilities of the \textsc{Fluka} code package, we compare \textsc{Fluka} predictions with neutron measurements at different altitudes, see \cref{fig:1}.

For comparisons with experimental data presented in the following, first the local interstellar spectrum of Galactic cosmic rays has been generated according to the \textsc{Bess} and \textsc{Ams} experiments, accounting for the modulation by the solar activity using the model in Ref.~\cite{SOLMOD} according to the counting rate of the \textsc{Climax} neutron monitor for the period of the experimental measurements.

The primary particles, from protons to nickel ions, are then injected at the top of the atmosphere, at an altitude of about 100\,km.
The geomagnetic cutoff is then applied using the Stormer formula for a Earth-centered magnetic dipole, with the strength adapted so that it reproduces the vertical cutoff specific to the geographical location where the measurements took place.
The maximum sampling energy is 100\,TeV, more than sufficient to assure that the disregarded high-energy part of the cosmic-ray spectrum will not give any detectable contribution.
The electromagnetic component of air showers is neglected in this case, since the resulting photoproduced neutrons are in negligible numbers given that most of the neutrons originate from showers initiated below 100\,GeV.

Two sets of experimental data have been used, the measurements taken aboard an ER-2 airplane at high altitude in 1997 and subsequently at ground level in 1999~\cite{Ref13}, and those taken on top of the Zugspitze mountain (2963\,m) in 1995~\cite{Ref14}.
Both experiments measured the neutron intensity and spectra using a Bonner multi-sphere spectrometer.
As such the reported spectra were obtained by unfolding the count readings of the spheres, a procedure which is somewhat dependent on the guess spectrum used for driving the unfolding procedure, and on the accuracy of the computed response functions of the Bonner spheres.
On this last point, it should be stressed that in the former experiment, there were only two spheres sensitive above a few tens of MeV, in the latter only one, and in both cases their response functions were taken from simulations, rather than being based on experimental or evaluated cross sections.
As a consequence, the spectral features above ${\sim}20$\,MeV are those coming from the guess spectra, and the integral intensity of the high energy component is dependent on the accuracy of the simulations used for the computation of the response functions.
A more sound approach, completely independent of the unfolding assumptions, would have been to compare directly the count rates of the various spheres with those computed using the same code for the air shower and detector response calculations, as for example was done in Refs.~\cite{Birattari:1994bm,Birattari:1997gc} in previous validations of \textsc{Fluka} predictions around high energy accelerators.
Unfortunately, neither experiment reported directly the count rates of the Bonner spheres.

Another source of uncertainty arises from the limited knowledge of the precise conditions in which the data were taken. 
In particular, the air humidity or lack thereof and the ground composition and water content can greatly affect the shape and intensity of the low energy part of the neutron spectra.
Also, for what concerns the data taken in flight, the influence of the aircraft material surrounding the Bonner spectrometer can introduce significant uncertainties (see Ref.~\cite{PELLIC2} for example).
Sizable corrections for this effect were applied in the experimental analysis, however they were again based on calculations with codes, and not on any experimental finding.
For the data at the ground level of Ref.~\cite{Ref13}, concrete as ground material and 50\% air humidity was assumed, while for the data of Ref.~\cite{Ref14} a mix of soil and snow and saturated air humidity was taken, knowing that during the measurements it was raining and there was snow on the ground.

Given these limitations, the measured data are well reproduced by the \textsc{Fluka} simulations.

Another benchmark with respect to neutron measurements in the atmosphere is presented in the following. 
In Ref.~\cite{Narita}, the count rates of 6 different neutron counters as measured on 27 February 1985, at two altitudes, are presented in \cref{tab:narita} and compared with \textsc{Fluka} predictions.
The neutron detectors comprised five Bonner spheres with polyethylene radii ranging from 4.1 to 22.6\,cm with a 10\,bar $^3$He counter in the centre, plus the bare $^3$He counter itself.
The latter is mostly sensitive to thermal neutrons only, while the response function of the spheres is progressively moving to higher neutron energies with the increasing radius.
The advantage of this comparison is that it allows us to use the same code for estimating both the detector response functions, and the neutron fluences in the atmosphere, and then compare directly the measured and calculated count rates for each detector.
In this way all uncertainties, connected with using response functions computed with other codes and with the unfolding procedure, are eliminated.
Of course, the systematic uncertainties, related to the actual atmospheric conditions (pressure and humidity) on the day of the measurements and on the effect of the aircraft structure, cannot be eliminated.
Those uncertainties are mostly affecting the expected thermal neutron component, and therefore the count rates of the bare $^3$He counter.
Good agreement between the \textsc{Fluka} predictions and the measurements is found.

\begin{table}
\caption{Altitude variation of multi-sphere neutron spectrometer rates: comparison between experimentally measured count rates~\cite{Narita} taken in flight on 27 February 1985 and \textsc{Fluka} results.
For the bare detector, which is sensitive to only almost-thermal neutrons, two values are given: assuming no aircraft skin and interior material ($^\dag$), and assuming a 4\,mm thick aluminium aircraft skin and a 6\,mm plastic equivalent interior material ($^\ddag$).}
\label{tab:narita}
\begin{center}
\begin{tabular}{ccccc}
\hline\hline
                   & \multicolumn{4}{c}{Altitude}
\\
\cline{2-5}
Poly               & \multicolumn{2}{c}{4880\,m}  
                   & \multicolumn{2}{c}{11\,280\,m}
\\
radius             & measured & \textsc{Fluka} 
                   & measured & \textsc{Fluka}
\\
(cm)               & $(s^{-1})$
                   & $(s^{-1})$
                   & $(s^{-1})$
                   & $(s^{-1})$
\\
\hline
\multirow{2}{*}{0}
                   & \multirow{2}{*}{$0.430\pm0.04$}
                   & $0.209^\dag\pm0.002$
                   & \multirow{2}{*}{$2.61\pm0.02$}
                   & $1.02^\dag\pm0.01$
\\
                   & 
                   & $0.500^\ddag\pm0.002$
                   &
                   & $2.45^\ddag\pm0.02$
\\
4.1                & $0.570\pm0.04$
                   & $0.591\pm0.005$
                   & $3.31\pm0.14$
                   & $2.90\pm0.02$
\\
5.6                & $0.753\pm0.05$
                   & $0.937\pm0.007$
                   & $4.50\pm0.16$
                   & $4.40\pm0.03$
\\
7.6                & $0.860\pm0.05$
                   & $1.010\pm0.01\phantom{0}$
                   & $5.01\pm0.17$
                   & $4.75\pm0.04$
\\
11.6               & $0.620\pm0.05$
                   & $0.648\pm0.004$
                   & $3.35\pm0.14$
                   & $3.07\pm0.02$
\\
22.6               & $0.177\pm0.02$
                   & $0.160\pm0.001$
                   & $1.09\pm0.08$
                   & $0.865\pm0.006$
\\
\hline\hline
\end{tabular}
\end{center}
\end{table}

\section{Simulation Method for Extensive Air Showers}
\label{sec:simmethod}

The calculations have been performed for vertical incidence, at five different primary energies in the energy range from $5.6{\times}10^{14}$ to $5.6{\times}10^{18}$\,eV, for photon, proton, and iron primaries, respectively.
In the shower evolution, hadrons and leptons have been simulated down to 5\,MeV, with unstable particles being allowed to decay, annihilate, or get captured, with the exception of neutrons which have been transported down to $10^{-5}$\,eV. 
All relevant physics processes have been simulated, including electromagnetic dissociation and photonuclear interactions.
Spectra of muons, neutrons, protons, charged pions, anti-protons, and anti-neutrons as representative particles have been recorded at 8 different atmospheric depths (196, 297, 399, 492, 594, 675, 878, and 1033\,g/cm$^2$) for several radial distances and time intervals.
The radial intervals are chosen to cover the features of the lateral distribution of showers at about $10^{16}$\,eV ranging from $(0<R/\text{m}<50)$ to $R>400\,$m in five intervals.

To capture the characteristic features of the arrival times, we use a logarithmic binning in time delays.
We define the arrival time of a particle as the delay with respect to the travel time of light from the first interaction point to the impact point with the measurement plane.
For muons, the binning is in factors of $10^{1/3}$, starting with $10^{-1}$\,ns.
Because of the different time scales relevant to neutrons, the binning for neutrons is starting with 1\,ns and extends until 10\,ms.

For the simulations shown here, the Earth's atmosphere has been described with 100 layers of increasing density, according to the US Standard Atmosphere, the same approach successfully used in the past in Refs.~\cite{Battistoni:2001fp,Zuccon:2003ns,Battistoni:1999at,PELLIC1,PELLIC2}.
The atmosphere has been assumed to be completely dry, and for the ground a typical soil composition with some level of moisture has been adopted. 
For two special cases we add a representative ground level to study the effect of the ground on the particle spectra.
Resembling the situation at the Pierre Auger Observatory~\cite{PierreAuger:2015eyc}, we add the same soil composition mentioned before as ground level at an atmospheric depth of 878\,g/cm$^2$, and a ground layer of ice at 675\,g/cm$^2$ to study the situation similar to \textsc{IceCube/IceTop}~\cite{IceCube:2012nn}.

To aid the interpretation of the obtained spectra, and to obtain a proxy for the longitudinal profile of the shower development, we store the physical track lengths $\ell$ of different particle types separately per each atmosphere layer.
With this information, we can calculate (per primary cosmic-ray) the fluence $f$ of a specific particle type in an atmospheric layer $i$ with density $\rho_i$ and thickness $\Delta X_i$ using the total track length $\ell_i$ summed over all particles $j$ of this type, $\ell_i=\sum_j\ell_{ij}$, as
\begin{equation}
  f(X_i) = \frac{\rho_i\,\ell_i}{\Delta X_i}. 
\label{eq:fluence}
\end{equation}
A proxy for the longitudinal profile is thus given by $f(X_i)$ as a function of the mean depth of the layer.
This fluence can be understood as following: By multiplying the track length with the density of the atmosphere, we transform this quantity into an effective grammage traversed by the particle tracks. 
Normalising this total track length to the thickness of the layer $\Delta_i$ in g/cm$^2$ results in a unit-less quantity that is approximately proportional to the particle fluence.

In the longitudinal shower profiles typically shown in air shower physics, the number of particles is counted when crossing virtual planes located at different depths (and usually oriented perpendicularly to the shower axis).
Differences between the longitudinal profile of fluence calculated here and the classical longitudinal shower profile arise mainly from the angular distribution of the shower particles with respect to the shower axis.
If all particles were propagating only parallel to the shower axis, the two quantities would be equal, up to the accuracy imposed by the binning of the atmospheric layers.
In fact, the fluence as described here is a better observable for low-energy particles in showers because particles traveling in all directions are considered equally.

\section{Results}
\label{sec:results}

In this section, we show energy spectra of particles obtained with our simulations.
We aim to highlight the characteristic features of neutrons in air showers by comparing their spectra to those of more well-known shower particles such as muons.
We consider different primary particles and also investigate the contribution of different processes to the overall neutron production.
Only a selection of the results will be shown here to illustrate the most important features on a qualitative level even though the simulations are producing numerically accurate results.
As discussed in \cref{subsec:flukagcr}, the dependence of neutron spectra on various environmental conditions will require detailed simulations for concrete setup to obtain more quantitative predictions at low energies.

The energy spectra shown in the following are normalized such that, if integrated, the number of secondary particles (arriving at the ground) per shower is obtained.
In general, the statistical errors on the individual bins are below 1\% and we chose not to show error bars for the sake of clarity of the presentation.
It should be noted that in sparsely populated regions of the phase space the statistical errors are larger than 1\% even when no error bars are shown.

We structure this discussion in sections to highlight the expected scaling of hadronically produced particles with depth, energy, and primary particle.
In addition, we will discuss arrival times and the effect of the ground on neutrons in air showers.

\subsection{Energy Spectrum and Production Mechanisms of Neutrons}

\begin{figure*}
\centering
\def\h{0.65}
\includegraphics[height=\h\textwidth]{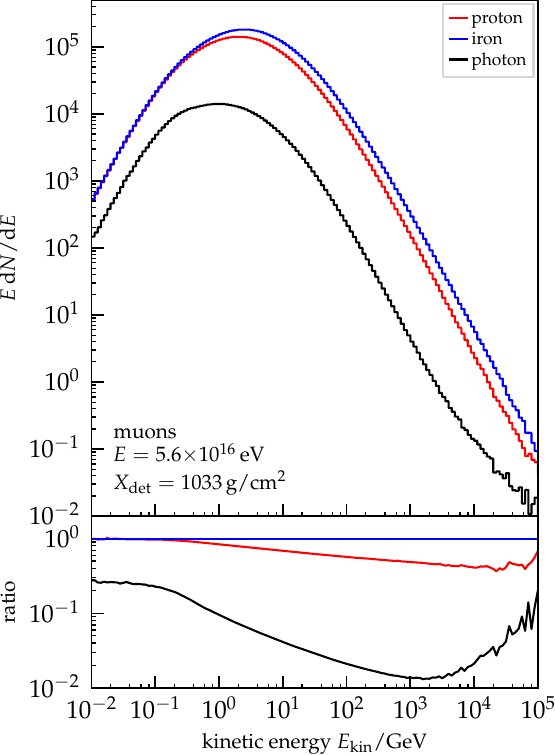}\hfill
\includegraphics[height=\h\textwidth]{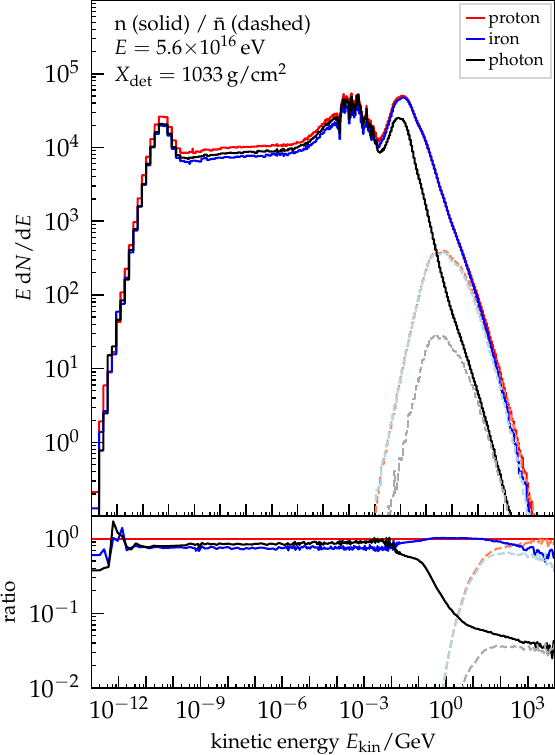}
\caption{\emph{Left:} Energy spectrum of muons in simulations performed with different primaries.
The energy $5.6{\times}10^{16}$\,eV is identical for all primaries and the muon spectrum is calculated at ground level.
The lower panel shows the ratio of the spectra to the iron spectrum, highlighting the expected scaling of the muon number with primary mass.
\emph{Right:} Energy spectrum of neutrons (solid) and anti-neutrons (dashed) in simulations of different primaries with energy $5.6{\times}10^{16}$\,eV at ground level.
The lower panel shows the ratio of the spectra to the proton spectrum.
For lower neutron energies a universal spectrum is observed, common to both hadronic and electromagnetic showers.
At higher neutron energies a clear difference between photon and hadron showers is visible.
The comparison with the anti-neutron spectra highlights hadronic pair-creation as the main production mechanism at those neutron energies.
Proton and iron showers are remarkably similar in neutron spectra at these energies on the ground.}
\label{fig:mu_primarydep} 
\label{fig:neu_primarydep} 
\end{figure*}

The energy spectra obtained for muons and neutrons in showers of different hadronic primaries and photons with energy $E = 5.6{\times}10^{16}\,$eV are shown in~\cref{fig:mu_primarydep}\,(left).
The energy spectra are shown for particles crossing a virtual plane at depth 1033\,g/cm$^2$, which corresponds to sea level.

The energy spectra of muons exhibit the characteristic depletion below 1\,GeV due to muon decay and energy loss in the atmosphere, which is dominated by ionization energy loss at low energies.
The importance of hadronic interactions feeding the main muon production channels (pion and kaon decay) is clearly visible with the muon number being smaller by more than a factor of 10 in photon-induced showers.
Also, as expected in the superposition~\cite{Engel:1992vf} and the Heitler-Matthews models~\cite{Matthews:2005sd}, see also Ref.~\cite{Gaisser:2016uoy}, the number of high-energy (approx.\ $E_\upmu > 1$\,GeV) muons is larger for iron showers than for proton showers.

In comparison, the neutron spectra, shown in \cref{fig:neu_primarydep}\,(right), for the same showers extend to much lower energies.
Because the neutron life time is much larger than the typical time scales of air showers and because of the absence of ionization losses, neutrons populate the full energy spectrum down to thermal energies.
The high-energy part of the neutron spectrum is fed by neutrons produced in the same hadronic interactions as pions and kaons, which give rise to muons, except that the neutrons are pair-produced with anti-baryons due to baryon-number conservation.
The neutrons of the primary nucleus -- like iron -- are negligible in number.
We can see this effect confirmed by observing the anti-neutrons (dashed) in~\cref{fig:neu_primarydep}\,(right) that perfectly follow the neutron spectra above energies of several GeV.
Confirming this explanation of hadronic interactions feeding the high-energy end of the spectrum, the number of high-energy neutrons in photon showers is much lower, as can be easily seen in both the absolute number -- \cref{fig:neu_primarydep}\,(right) top -- or the ratio distribution in the lower panel.

The spectra below a few GeV is showing more features that originate from different processes.
Intermediate and low-energy neutrons are produced in intra-nuclear cascade re-interactions and in nuclear de-excitation processes. 
The latter, when occurring in a projectile nucleus, will results in the most energetic nucleons when transformed to the laboratory frame.

\begin{figure}
    \centering
    \includegraphics[width=\columnwidth]{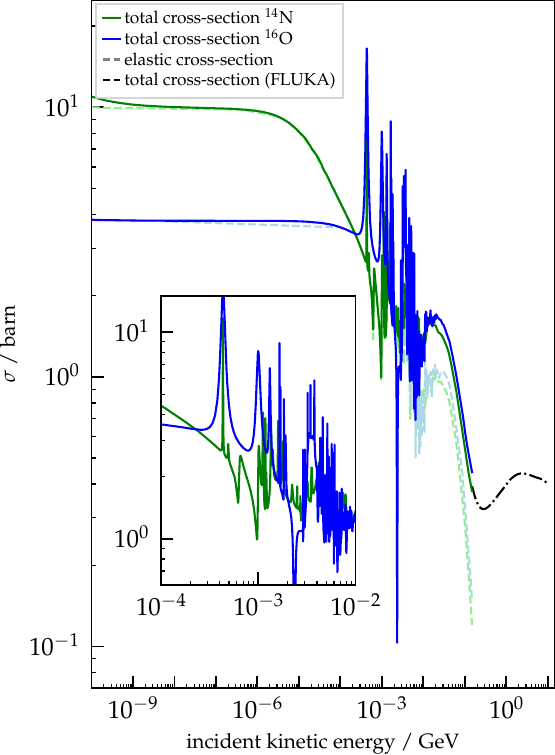}
    \caption{Cross sections of neutrons for different target nuclei representing the atmosphere.
    The inset is highlighting the resonance structure at around 1\,MeV of neutron kinetic energy.
    Dashed colored lines indicate the elastic cross section for the two chosen target nuclei.
    The cross-section data have been taken from the evaluation \textsc{Endf/B-Viii.0}~\cite{Endfb8r0}. 
    Above 150\,MeV, we show the total cross section from \textsc{Fluka}~\cite{Prael:1999} for nitrogen as dash-dotted black line to highlight features at higher energies.}
    \label{fig:cross_sections}
\end{figure}

The broad feature in the neutron spectrum in the region from 50 to 150\,MeV is referred to as the quasi-elastic peak.
It is the result of the combination of non-elastic neutron-nucleus interactions made up by neutron-nucleon elastic interactions below the pion production threshold and the broad minimum in nucleon-nucleon cross sections between 100 and 300\,MeV. 
The additional structures visible between about 0.1 to 10\,MeV are the result of many resonant cross-section channels that depend on the target material.
As typical examples in air, we show this resonance structure for the total and elastic cross-sections of neutrons on nitrogen and oxygen in \cref{fig:cross_sections}.

It should be noted that the cross section for photo-nuclear interactions is maximal in the delta-resonance region (around 150 to 450\,MeV), and in the Giant Dipole Resonance (GDR) region (around 10 to 40\,MeV for light nuclei).
In the delta-resonance region, low-energy pions and nucleons spanning up to a couple of hundreds MeV can be produced, in the GDR region mostly neutrons of a few MeV are produced.
Therefore, photo-nuclear interactions are expected to contribute more to the low-energy part of the neutron spectrum.
This expectation is clearly on display in \cref{fig:neu_primarydep}, particularly when looking at the bottom panel displaying the ratio of the spectra.

Finally, below 10 to 20\,MeV, the neutron propagation is best described as diffusion in the atmosphere.
As a consequence, the angular distribution of these neutrons with respect to the shower axis is very wide. 
In an ideal scenario, for a perfect moderator with negligible absorption, the energy spectrum below this resonance region follows an $\dd N/\dd E\propto E^{-1}$ spectrum.

To understand this result, we have to consider the kinematics of elastic scattering of neutrons on nuclei.
From a classical elastic scattering, we find the minimal kinetic energy $E^\text{min}_\text{kin, n}$ of the scattered neutron as
\begin{equation}
E^\text{min}_\text{kin, n} =
  E_\text{kin,n} \left(\frac{M_\text{t} - m_\text{n}}{M_\text{t} + m_\text{n}}\right)^2 \equiv
  \alpha \, E_\text{kin,n}
\label{eq:elastic_scattering}
\end{equation}
with the neutron mass $m_\text{n}$ and the mass of the target nucleus $M_\text{t}$.
For a derivation, see \cref{app:elastic_scattering}.

At the energies we are interested in and below the threshold for inelastic scattering, the elastic scattering distribution is isotropic in the center-of-mass system
\begin{equation}
\frac{\dd N}{\dd\Omega^\ast} \sim \frac{\dd N}{\dd\cos\theta^\ast} \sim \text{const.}
\end{equation}
where $\theta^\ast$ denotes the scattering angle.
Using the transformation to the laboratory system
\begin{equation}
E = \gamma E^\ast + \gamma\beta p^\ast \cos\theta^\ast
\end{equation}
we get $\dd E \sim \dd\cos\theta^\ast$ and, consequently,
\begin{equation}
\frac{\dd N}{\dd E} \sim \text{const.}
\end{equation}
Therefore, the probability for a neutron of initial energy $E'$ to emerge from an elastic scattering with an energy between $E$ and $E+\dd E$ is given by
\begin{equation}
P(E'{\to}E) \, \dd E =
\begin{cases}
\frac{1}{(1-\alpha)E'}\,\dd E & :\alpha E' < E \leqslant E',
\\
0 & :E \leqslant \alpha E'.
\end{cases}
\label{eq:scattprob}
\end{equation}
Hence, the propagation of neutrons through the atmosphere leads to a shift of their energies to ever lower and lower values, building up a low-energy tail of the neutron distribution.
This tail grows until a neutron energy distribution close to $\Phi(E)=C/E$ is reached.
For each further propagation step through an atmospheric depth layer of thickness $\dd X$, the energy spectrum of the neutrons changes by
\begin{equation}
\frac{\dd\Phi(E)}{\dd X} =
  -\frac{\sigma(E)}{M_\text{t}} \Phi(E) +
  \int_E^{\frac{E}{\alpha}} \frac{\sigma(E')}{M_\text{t}}\,P(E'{\to}E) \, \Phi(E') \, \dd E',
\label{eq:balance}
\end{equation}
where the first term on the r.h.s.\ describes the loss of neutrons of energy $E$ undergoing an interaction in this depth layer.
The second term accounts for neutrons of higher energy $E'$ that are scattered to an energy in the interval $[E, E+\dd E]$.
For a neutron flux $\Phi(E)\propto E^{-1}$, both terms have an energy dependence of $\propto E^{-1}$ if the cross section $\sigma(E)$ for elastic scattering is energy-independent (which is the case in the considered energy range).
Therefore, in the limit $\Phi(E)\to C/E$, only the normalization but not the energy spectrum of the neutrons changes due to the propagation through atmosphere.
Indeed, the results of the simulations show that neutrons exhibit to a good approximation a $\dd N/\dd E\propto E^{-1}$ spectrum at these energies over a wide range of atmospheric depths.

The lower end of the neutron spectrum is given by the thermal peak that is prominently visible only if the target material has sufficient hydrogen to fully thermalize the neutrons -- a condition in this case fulfilled only for the ground level because of the simulated soil at 1033\,g/cm$^2$.
This feature is visible in \cref{fig:neu_primarydep}\,(right) as thermal peak in the neutron energy spectrum.

While the expected difference in muon number for different primary particles is reproduced in our simulations and is best visible in the lower ratio panel of \cref{fig:mu_primarydep}\,(left), the neutron spectra of protons and iron nuclei are very similar up to ${\sim}10$\,GeV, see \cref{fig:mu_primarydep}\,(right).
The comparison with anti-neutrons shows that the low-energy part of the neutron spectrum is originating from neutrons released in the disintegration of nuclei of the atmosphere.
This disintegration is caused by both the interaction of hadrons and the interaction of the much more abundant photons with air nuclei in an air shower.
The equality of neutrons from proton and iron showers is a numerical coincidence governed by the shower development that we discuss in the next Section.

\subsection{Depth Evolution}

\begin{figure*}
\centering\def\h{0.65}
\includegraphics[height=\h\textwidth]{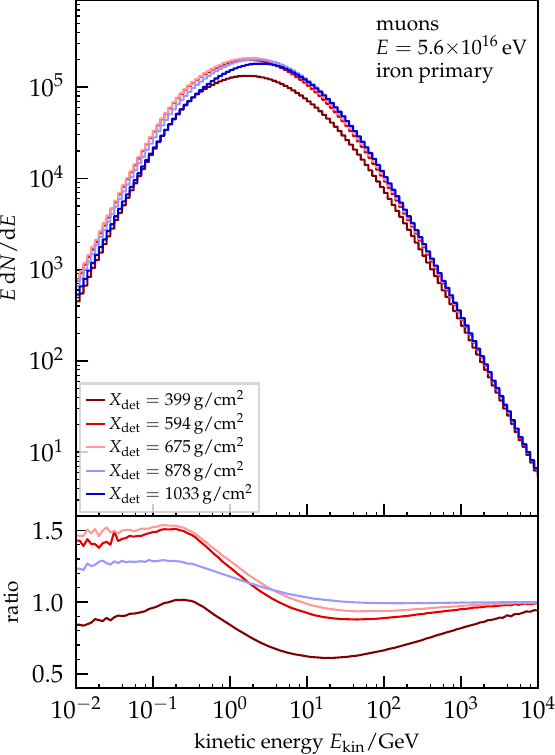}\hfill
\includegraphics[height=\h\textwidth]{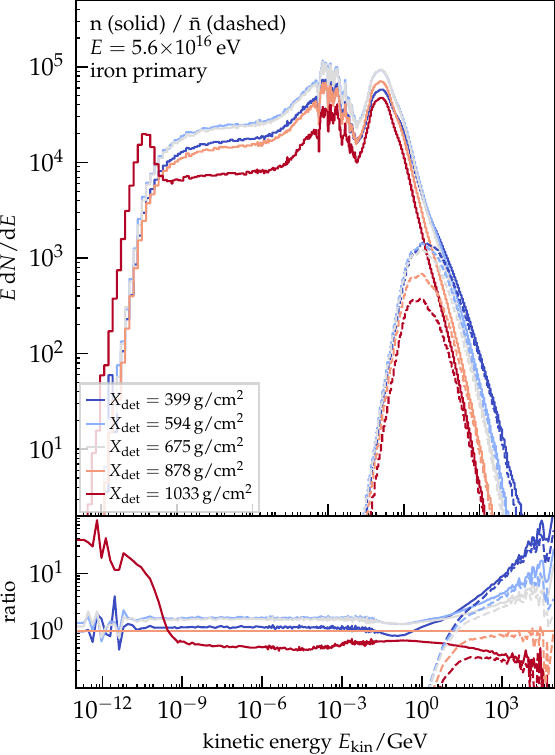}
\caption{
Comparison of muon and neutron spectra for different atmospheric depths for iron showers at $5.6{\times}10^{16}$\,eV.
\emph{Left:} Differences of the muon energy spectra at different atmospheric depths.
The lower panel shows the ratio of the spectra to the ground-level one at $X_\text{det}=1033\,$g/cm$^2$ to increase the visibility of the changes.
\emph{Right:} Energy spectrum of neutrons (solid) and anti-neutrons (dashed) in simulations of $5.6{\times}10^{16}$\,eV iron showers at different atmospheric depths.
The lower panel shows the ratio of the spectra to the 878\,g/cm$^2$ spectrum neutron spectrum on a logarithmic scale.}
\label{fig:muon_xdep}
\label{fig:neu_xdep}
\end{figure*}

Following the discussion of the energy spectra of neutrons and their production mechanisms in the previous section, we focus on the development of the neutron component of air showers with atmospheric depth to understand the numerical coincidence of equal neutron numbers for proton and iron showers.

Similarly to the previous discussion on the overall energy spectrum, we here also use the muon component as reference for comparison.
A first hint towards explaining the difference in neutron dependence on primary mass can be inferred from the energy spectra, as used in the previous section, but highlighting different atmospheric depths.
\cref{fig:muon_xdep} shows the energy spectra of muons (left) and neutrons (right) for a set of five different atmospheric depths bracketing the maximum of the shower.
From the distribution of the muons in the left panel, we can see -- with the exception of the spectrum at $X_\text{det}=399\,$g/cm$^2$ before the maximum -- that the variations of the high-energy muons are less than 20\%.
This expected result highlights the absence of strong attenuation of muons in the atmosphere.

In contrast, the right panel of \cref{fig:muon_xdep} shows the spectra of neutrons in the same atmospheric depths with strongly visible variations.
The bottom panel, displaying the ratio of the spectra in logarithmic scale, clearly emphasizes the different processes relevant for these strong differences.
At the lowest thermal energies, the effect of moderation in the ground dominates and leads to the aforementioned thermal peak of neutrons at $X_\text{det}=1033\,$g/cm$^2$.
For the highest energies, we can also see a clear difference to muons: the neutrons still interact in the atmosphere down to the ground level and thus shift their distribution from the highest energies to the MeV range.
In the diffusive regime with an $E^{-1}$ spectrum, we observe the evolution of the shower already in the particle abundance: a rise from 399\,g/cm$^2$ up to 675\,g/cm$^2$ with a subsequent decrease in overall numbers.
This decrease is indicative of the (expected) stronger attenuation that neutrons experience compared to muons.

\begin{figure*}
\centering
\def\h{0.65}
\includegraphics[height=\h\textwidth]{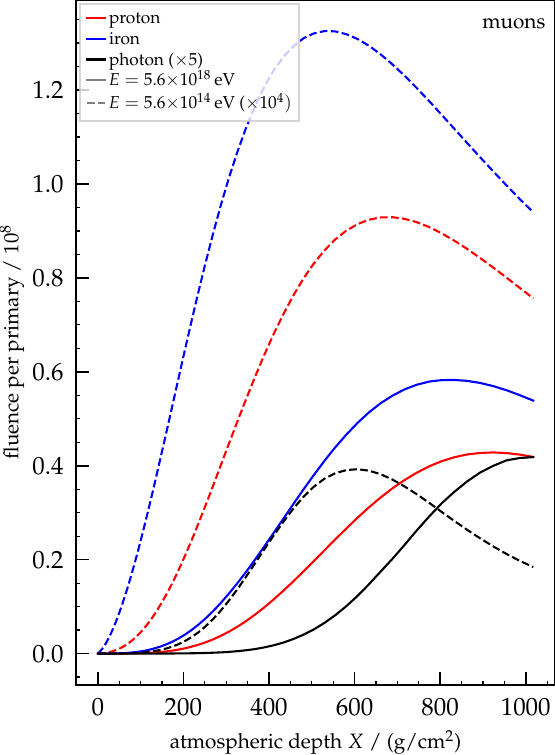}\hfill
\includegraphics[height=\h\textwidth]{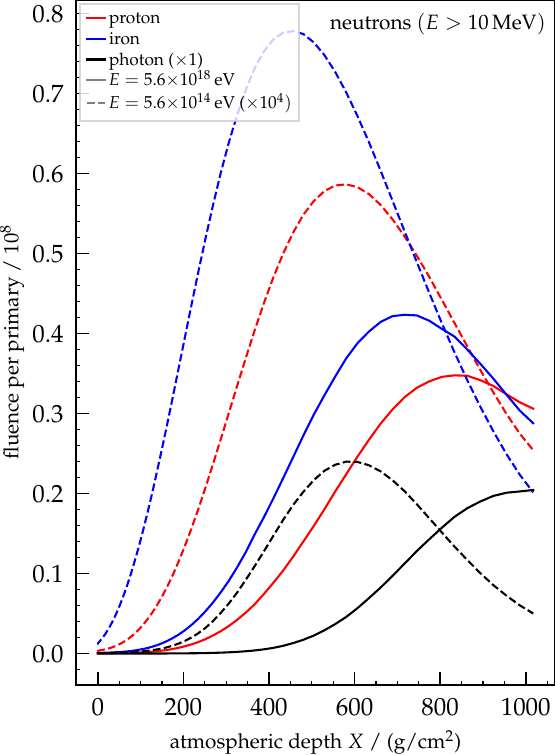}
\caption{Longitudinal profiles for proton (red), iron (blue), and photon (black) showers for two energies of $5.6{\times}10^{14}$\,eV (dashed) and $5.6{\times}10^{18}$\,eV (solid) scaled inversely linear with energy to highlight deviations from linear scaling.
We use the particle fluence as calculated in \cref{eq:fluence} for the profiles and show a linear interpolation between depths defined in \cref{sec:simmethod}.
\emph{Left:} Longitudinal profiles of muons, with photon-induced profiles scaled with a factor five for visibility reasons.
\emph{Right:} Longitudinal profiles of neutrons with kinetic energies above 10\,MeV.
We select only neutrons with kinetic energies above 10\,MeV in the profiles shown to highlight the parts potentially detectable in cosmic-ray experiments.
Note that the photon profiles are not multiplied with a factor five.}
\label{fig:profile_primary_muon} 
\label{fig:profile_primary_neu10MeV} 
\end{figure*}

To further highlight the difference in attenuation, we can use the longitudinal profiles constructed from \cref{eq:fluence}.
\cref{fig:profile_primary_muon} shows such profiles for muons and neutrons with kinetic energy above 10\,MeV for different primaries and showers of two energies.
We scale the profiles with the inverse of the energy to highlight deviations from scaling with $E$.
In the left panel, for muons, we see that after the shower maxima the profiles of proton and iron showers run almost parallel, approximately preserving the difference obtained at the maximum.
For neutrons in the right panel, as expected from the observation in the energy spectrum, we see that the attenuation is much stronger and decisive for the observed similarities between proton and iron showers.
For both energies, we find a similar difference between proton and iron in neutron number at maximum as for muons.
However, the strong attenuation of neutrons -- usually referred to as neutron-removal length of the order of 100\,g/cm$^2$ -- combined with the shifted maximum positions leads to a cross-over point between the two profiles that is followed by very similar abundances of neutrons in proton and iron showers at depths between 850 to 1000\,g/cm$^2$.

Thus, we can clearly identify the different attenuations of neutrons and muons as origin of the unexpected coincidence that proton and iron shower have very similar numbers of neutrons at the ground.
From the shift of the cross-over point between $5.6{\times}10^{14}$\,eV (about 750\,g/cm$^2$) and $5.6{\times}10^{18}$\,eV (about 950\,g/cm$^2$) we can also predict that the exact equality and observed numbers in non-vertical incidence might be very sensitive to the exact conditions of the atmosphere and the incidence angle.

\subsection{Energy Scaling of Particle Spectra}

Following the discussion on the shower development and the dependence of the neutron and muon spectra on atmospheric depth, in this section we analyse the scaling of these spectra with the primary energy.

\begin{figure}
\centering
\includegraphics[width=\columnwidth]{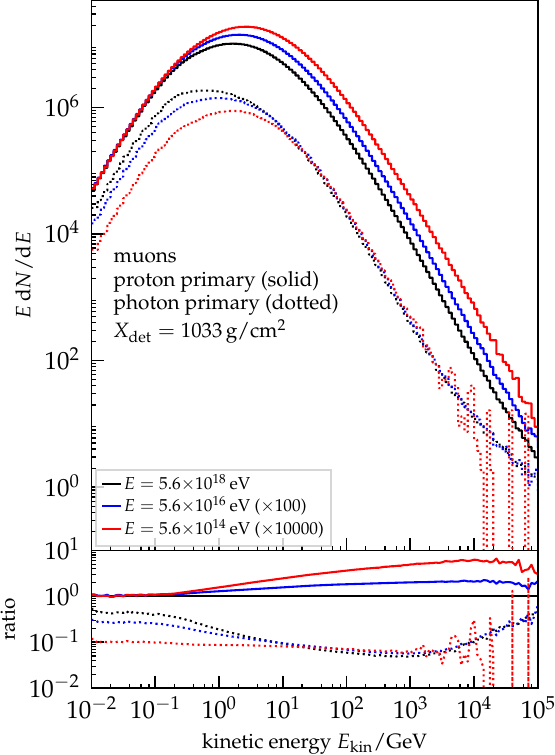}
\caption{Differences of the muon energy spectra for proton (solid) and photon (dotted) showers of different energies at ground level.
To take out the energy scaling, the spectra are multiplied with a factor following the shower energy $E^{-1}$.
The lower panel shows the ratio of the spectra to the proton spectrum at $E=5.6{\times}10^{18}$\,eV on a logarithmic scale.}
\label{fig:muon_edep}
\end{figure}

For hadronically produced particles, like muons, we expect from the Heitler-Matthews model~\cite{Matthews:2005sd} that their number increases roughly with $E^\beta$ with $\beta\approx0.94$.
In our simulations, we see that after re-scaling with the primary energy, the muon flux of the highest primary energy is lower than that of the lower primary energy, as shown in the spectra of \cref{fig:muon_edep}.

In contrast, muon production in photon showers scales linearly with energy, again in agreement with the expectations of the Heitler model, as evident from the high-energy tail of the muon spectra, shown as dashed lines in \cref{fig:muon_edep}.

\begin{figure}
\centering
\includegraphics[width=\columnwidth]{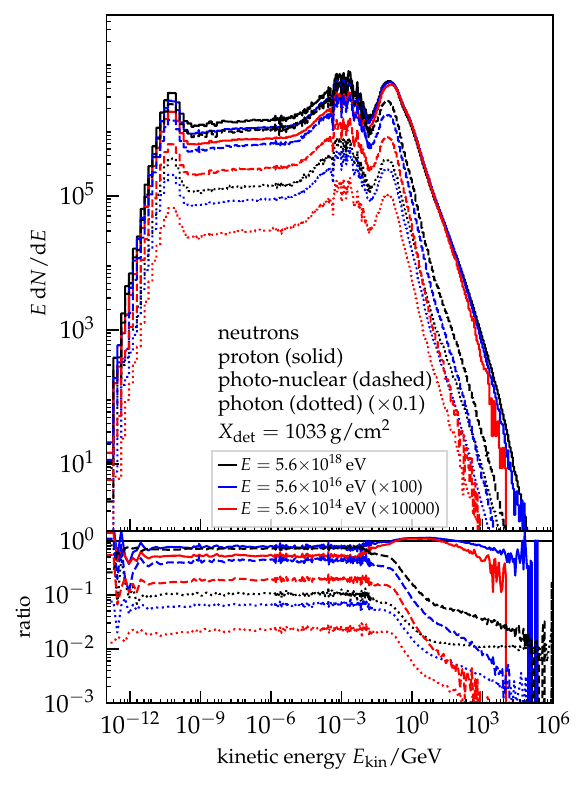}
\caption{Differences of the neutron energy spectra for proton (solid) and photon (dotted) showers of different energies at ground level.
Neutrons from photo-nuclear reactions in proton showers are shown as dotted lines for the different energies.
To take out the energy scaling, the spectra are multiplied with a factor following the shower energy $E^{-1}$.
To increase the visibility, spectra from photon showers are additionally scaled with $0.1$.
The lower panel shows the ratio of the spectra to the spectrum of proton showers at $E=5.6{\times}10^{18}$\,eV.}
\label{fig:neu_edep}
\end{figure}

Surprisingly, such a linear energy scaling seems also to be present in the case of high-energy neutrons, as \cref{fig:neu_edep} shows.
However, this is again the result of two effects.
Firstly, the number of neutrons increases with energy in the same way as those of muons. 
Secondly, the shift of higher-energy showers deeper into the atmosphere reduces the attenuation of neutrons and, hence, additionally increases the number of neutrons at the ground.
This attenuation effect is negligible for muons, as discussed in the previous section.

Effectively, a scaling of the neutron flux (at high energies only) approximately proportional to the shower energy is found for observation depths deeper than the shower maximum in our simulations.
With the shower maximum of vertical showers reaching the ground for typical observation sites at energies beyond $10^{18}$\,eV, this scaling is expected to be broken for energies beyond those of our simulations.

Similarly to the approximate coincidence of neutron numbers for iron and proton showers, we can also further analyse this surprising scaling with energy using the longitudinal profiles based on the fluence defined in \cref{eq:fluence}.

\begin{figure*}
\centering
\def\h{0.65}
\includegraphics[height=\h\textwidth]{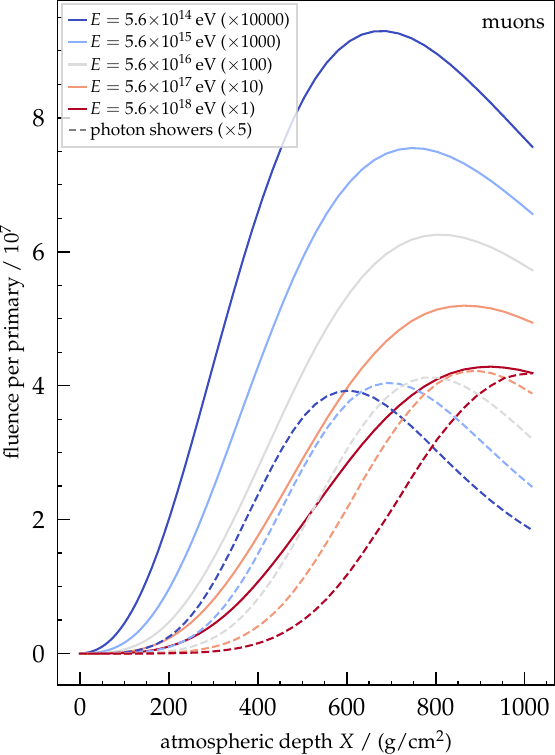}\hfill
\includegraphics[height=\h\textwidth]{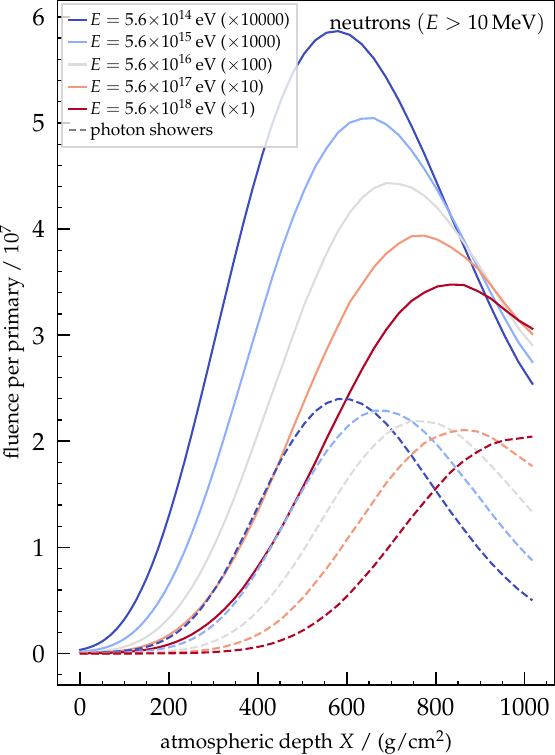}
\caption{
Longitudinal profiles for proton (solid) and photon (dashed) showers scaled inversely linear with energy to highlight deviations from linear scaling.
We use the particle fluence as calculated in \cref{eq:fluence} for the profiles and show a linear interpolation between depths defined in \cref{sec:simmethod}.
\emph{Left:} Longitudinal profiles of muons.
Photon shower profiles are additionally multiplied with a factor five to enhance visibility.
\emph{Right:} Longitudinal profiles of neutrons.
We select only neutrons with kinetic energies above 10\,MeV in the profiles shown to highlight the parts potentially detectable in cosmic-ray experiments.}
\label{fig:profile_energy_muon} 
\label{fig:profile_energy_neu10MeV} 
\end{figure*}

In \cref{fig:profile_energy_muon}\,(left) we show for reference the fluence profiles of muons for showers of different energies.
By scaling inversely linear with energy we can highlight the expected scaling at any depth.
With the exception of showers with an energy of $E=5.6{\times}10^{14}\,$eV, the profiles are parallel for different energies and follow the expected scaling.
It is also clearly visible that for photon showers -- dashed lines rescaled with an additional factor 5 for visibility -- the linear scaling is true, if losses due to attenuation are ignored.
For practical purposes this is a good approximation given that the attenuation is of the order of 50\% over four decades in energy.

For neutrons, the situation is different, as already discussed in the previous sections.
In the right panel of \cref{fig:profile_energy_neu10MeV} we show the fluence profiles of neutrons with kinetic energies above 10\,MeV rescaled linearly in inverse shower energy.
We recover the expected scaling with energy -- the same as for muons in \cref{fig:profile_energy_muon} -- if we focus on the maximal number of particles.
The strong attenuation is clearly visible and the approximate linear scaling at a depth of 850\,g/cm$^2$ is recovered as cross-over point of the profiles of different energies.

Interestingly, contrary to muons, the number of neutrons produced in photon showers is closer to that in hadron showers, visible by the absence of additional scaling factors for the photon spectra in \cref{fig:profile_energy_neu10MeV}.
At sea-level and for $E=5.6{\times}10^{18}$\,eV, we expect only a 25\% decrease in neutron numbers compared to iron showers.
At different atmospheric depths this difference is enhanced due to the differences in $X_\text{max}$ of photon and hadronic showers.

Finally, we note the somewhat different energy scaling of the number of neutrons at maximum in comparison to the number muons at maximum.
The number of neutrons increases faster with shower energy than the number of muons.

All these differences can be qualitatively understood by considering the relative importance of hadrons vs.\ photons in the production processes of neutrons and muons.
While muons are almost exclusively produced in interactions of hadrons (only about 10\% of the muons are electromagnetically pair-produced), photons contribute about 40\% of the neutrons through electromagnetic dissociation of nuclei.
Hadronic interactions scale roughly with $E^{0.94}$.
In contrast, the photon contribution increases approximately linearly with the shower energy.
The production of neutrons by electromagnetic dissociation also explains the large number of neutrons found in photon showers.

\subsection{Arrival Times of Particles on Ground}

A characteristic feature of neutrons in air showers that has been used in earlier works~\cite{PhysRev.75.1532, Linsley:1984bnz} for their identification, is the arrival-time distribution.
Compared with other shower particles, many neutrons arrive ``late'' at the ground.
To be more quantitative, we want to measure the delay of a particle from the shower arriving at a given point on the ground.
We define the reference time from which the delay is calculated as the time after the first causal signal from the first interaction of the shower arrives at the given point, i.e.\ we measure the delay with respect to a virtual particle moving with the speed of light from the first interaction point to the location of observation.
As presented in \cref{sec:simmethod}, this information is available in the form of logarithmically time-binned energy spectra of particles.

\begin{figure}
\centering
\includegraphics[width=\columnwidth]{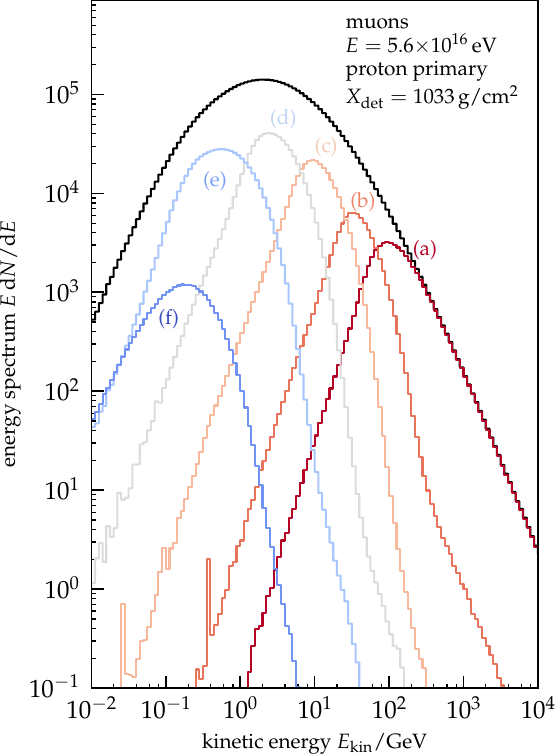}
\caption{Energy spectra of muons for different arrival times.
We show selected logarithmically spaced time bins for $5.6{\times}10^{16}$\,eV proton showers at ground level: $\text{(a)}=[0.1,0.215]$\,ns, $\text{(b)}=[0.464,1]$\,ns, $\text{(c)}=[4.64,10]$\,ns, $\text{(d)}=[46.4,100]$\,ns, $\text{(e)}=[464,1000]$\,ns, and $\text{(f)}=[4.64,46.4])\,\upmu$s.
As expected, the muons of highest energies have the smallest delays and overall arrival times are within a couple of microseconds for all muons.}
\label{fig:muon_tdep}
\end{figure}

To highlight our observations, we focus on the abundances of particles within given time intervals for showers from protons with $E=5.6{\times}10^{16}\,$eV at ground level.
The expected qualitative differences in these time distributions are comparably small for different primaries and energies.
In \cref{fig:muon_tdep} we show the energy spectrum of muons arriving in six different time windows spanning sub-nanosecond (red) to about 50\,$\upmu$s (blue) in delay.
As expected, we find that the fraction of muons arriving with delays of more than a few microseconds (bin (f) in \cref{fig:muon_tdep}) is very small.
The majority of the muons arrives with delays of around 100\,ns to several hundreds of nanoseconds, as is clearly visible from the peaking of the bin (d) in the distribution.
As expected, the highest energy muons have the smallest delay.

\begin{figure}
\centering
\includegraphics[width=\columnwidth]{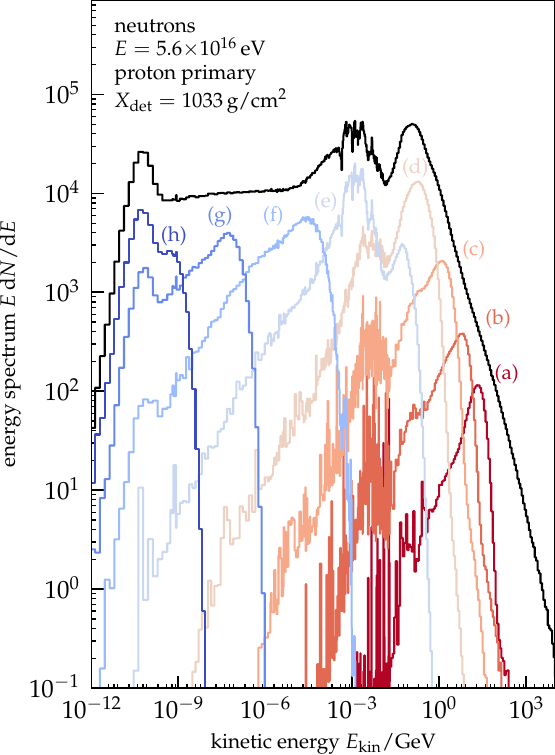}
\caption{Energy spectra of neutrons for different arrival times.
We show only selected time bins for $5.6{\times}10^{16}$\,eV proton showers at ground level:
$\text{(a)}=[4.64,10]$\,ns, $\text{(b)}=[46.4,100]$\,ns, $\text{(c)}=[464,1000]$\,ns, $\text{(d)}=[4.64,10]\,\upmu$s, $\text{(e)}=[46.4,100]\,\upmu$s, $\text{(f)}=[464,1000]\,\upmu$s, $\text{(g)}=[4.64,10]$\,ms, and $\text{(h)}=[46.4,100]$\,ms.}
\label{fig:neu_tdep}
\end{figure}

In contrast to the small delays of muons, the neutrons arriving at the ground span many orders of magnitude in their delay, as is visible in \cref{fig:neu_tdep}.
Note that the binning is different from \cref{fig:muon_tdep} and covers time delays from 4.6\,ns (red) to 100\,ms (blue).
For the bulk of the thermal neutrons, the arrival time is of the order of 0.1\,s while the highest energy particles have delays similar to those of muons.
However, the characteristic time scales for neutrons that can be expected to produce a measureable signal in a detector typically used in air shower arrays -- i.e.\ kinetic energies of several MeV up to 1\,GeV -- are 1 to 10\,$\upmu$s.
Thus, it depends on the exact energy relevant for detection and the applied data-taking scheme whether or not these neutrons are recorded in cosmic-ray events.
If we take the distributions (d) and (e) in \cref{fig:neu_tdep}, we see that depending on whether or not the detection sensitivity window is closer to the resonance region at a few MeV or to the quasi-elastic peak at a few hundred MeV, the typical time delay shifts by an order of magnitude from tens of microseconds to a few microseconds.
We compare the abundances of neutrons and muons in selected time windows targeting typical detection ranges in \cref{sub:detection} where we also briefly discuss the influence of the radial selection on the distribution of arrival times.

\subsection{Radial Distributions}
\label{sub:radial_distribution}

\begin{figure}
\centering
\includegraphics[width=\columnwidth]{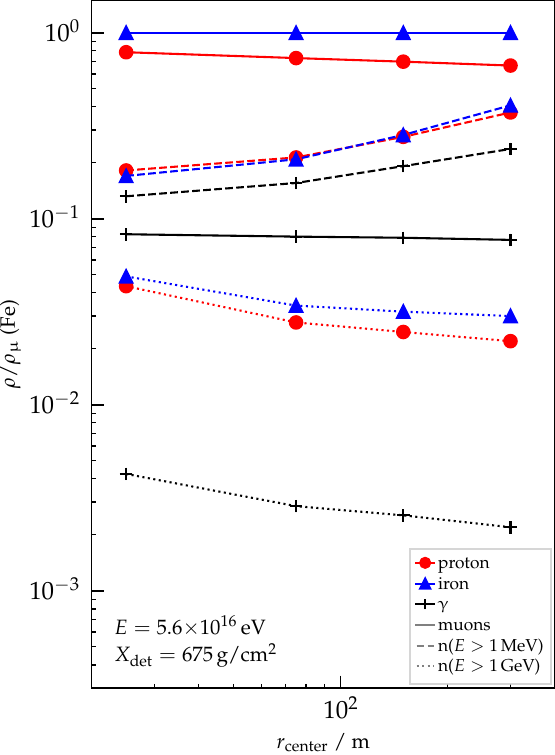}
\caption{Radial dependence of the muon and neutron distributions for different primaries.
Proton showers are shown in red, iron in blue, and photon showers in black.
We integrate muons over the full energy range (solid), and show two energy selections for neutrons (dashed, dotted).
The binning is the same as presented in \cref{sec:simmethod} and we use the arithmetic bin centers for plotting.
To increase the visibility of the differences we show the ratio of the respective density to the desity of muons in an iron shower $\rho(\text{Fe})$.
For the absolute particle density see the Appendix.}
\label{fig:ldf_like_ratio}
\end{figure}

Following the discussions of the energy spectra, we can use the information available in the radial bins of the simulations to derive features of the lateral distribution.
To obtain a lateral distribution, we integrate the energy spectra above a given minimal energy and divide it by the area of the radial interval $[r_1, r_2)$.
With the energy spectrum bins $\dd N(E_i)/\dd E$ of width $\Delta E_i$ above this minimal energy, we can write this as
\begin{equation}
\rho(r_\text{center}) = \frac{1}{\pi (r_2^2 - r_1^2)}\sum_i \Delta E_i \frac{\dd N(E_i | r1, r2)}{\dd E},
\end{equation}
where we have chosen the center of the lateral interval, $r_\text{center}{=}(r_2 + r_1)/2$, as reference point for plotting. 
In \cref{fig:ldf_like_ratio}, we show the resulting particle densities relative to the muon density in iron showers as function of the average radius $r_\text{center}$ of the bins for different primary particles at $E=5.6{\times}10^{16}$\,eV and $X_\text{det}=675\,$g/cm$^2$.
We chose the relative measure $\rho/\rho_\upmu(\text{Fe})$ to highlight differences between neutron and muon lateral distributions that in absolute numbers are dominated by the steeply falling overall lateral distributions.
For reference, we show the absolute lateral distributions in the Appendix.

Using neutrons above two different energies, 1\,MeV and 1\,GeV, we distinguish the two expected regimes of the lateral distributions of neutrons.
For high-energy neutrons above 1\,GeV, shown as dotted lines in \cref{fig:ldf_like_ratio}, the slope of the relative lateral density is negative, indicating that the high-energy neutrons are concentrated close to the shower axis, even more than muons.
To represent the lower-energy neutrons that could still be detected in cosmic-ray experiments, we use a minimum kinetic energy of 1\,MeV represented with dashed lines in \cref{fig:ldf_like_ratio}.
For these lower-energy neutrons, the slope is positive indicating, as expected, a wider distribution than for muons.

To quantify the abundance of neutrons in phase-space regions favourable to neutrons, we use the last available radial bin in \cref{fig:ldf_like_ratio}, $200\leq r/\text{m}<400$, and the lower energy threshold.
We find that, at $\rho_\text{n}/\rho_\upmu(\text{Fe})=41\%$, the number of neutrons is roughly half the number of muons for iron showers.
If we further apply a time selection of $[1,21.5]\,\upmu$s as typically experimentally necessary to distinguish neutrons from muons, we can further enhance the relative strength of the neutron number.
We obtain a relative density of 21\% for neutrons above 1\,MeV, while only 4\% of the muons arrive in this time interval.
Obviously, in real applications also the electromagnetic component has to be taken into account as well as details of the detector to define the minimal neutron kinetic energy.
Nonetheless, it is clear that phase-space regions with significant contributions of neutrons -- at least by particle number -- exist independent of the detailed assumptions.

In general, differences in the slope of the radial distributions of muons are expected for different primaries with steeper distributions for proton showers~\cite{Medina:2006my}.
We recover these differences in our simulations as visible in \cref{fig:ldf_like_ratio}.
Comparing the slopes for the neutrons, we observe the same behaviour for the high-energy component above 1\,GeV.
For the lower energy neutrons, the difference in slope vanishes.
From the discussion of the neutron energy spectrum we know that the electromagnetic component with photo-nuclear reactions is important for the neutrons at 1\,MeV.
Given the small differences in electromagnetic components of the proton and iron showers, this lack of difference is to be expected.
Similarly, we see that the lateral distribution of neutrons above kinetic energy of 1\,MeV in photon and hadron showers is rather similar.
On the contrary, at higher neutron energies, the neutron numbers are suppressed as already previously discussed.

\begin{figure}
\centering
\includegraphics[width=\columnwidth]{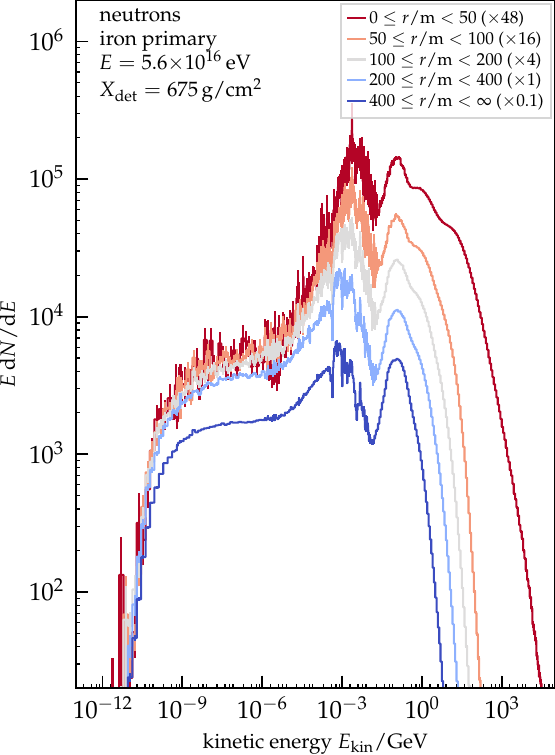}
\caption{Illustration of the different energy spectra of neutrons in different radial bins.
We show the results for iron showers at $E=5.6{\times}10^{16}$\,eV at an atmospheric depth of $X_\text{det}=675\,$g/cm$^2$.
The spectra in the radial bins are scaled with the inverse of their area.
The largest, unbounded interval is scaled with an approximate factor because no area can be defined.}
\label{fig:rdep_neu}
\end{figure}

We can also compare the energy spectra of neutrons in the different radial bins, as shown in \cref{fig:rdep_neu}.
The energy spectra shown are scaled with the inverse of their area relative to the largest bounded interval $200\leqslant r/\text{m}<400$.
The unbounded outer interval is multiplied with an approximate factor because no area can be analytically defined.
In the diffusive energy range of neutrons below several MeV, we can clearly see that the densities are similar following the expected diffusive behaviour up to almost 400\,m.
The other interesting feature visible in \cref{fig:rdep_neu} is the concentration of the highest energy neutrons around the shower axis.
While overall the neutron spectrum is softer away from the shower axis, there is an additional shoulder of neutrons around 30\,GeV that is only present in the interval up to 50\,m from the axis.

\subsection{Expected Neutron Densities and Prospects of Detection}
\label{sub:detection}

To illustrate the potential importance of neutron signals in air-shower measurements, we want to compare the densities of muons and neutrons for typical time windows recorded in existing observatories.
Furthermore, we highlight the effects the ground has on the energy spectra by using simulations with and without ground level at the observation depth.

\begin{figure}
\centering
\includegraphics[width=\columnwidth]{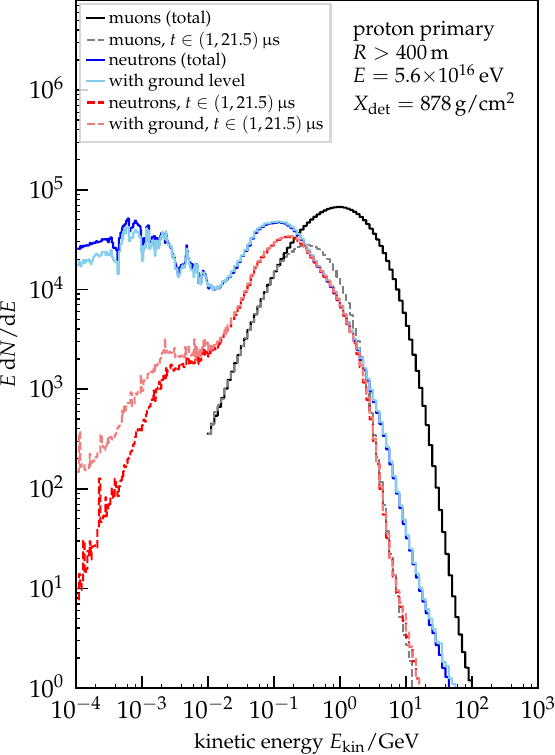}
\caption{Comparison of the energy spectra of muons and neutrons at an atmospheric depth of 878\,g/cm$^2$ for infinite times and typical cosmic-ray-event time scales.
We use proton showers at $5.6{\times}10^{16}$\,eV and distances greater than 400\,m from the core in this comparison.
Lighter colors indicate a dedicated simulation run with a ground level at this altitude, similar to the depth of the Pierre Auger Observatory.
We can infer that the neutron spectra are only influenced below a few MeV by the ground.}
\label{fig:auger_ground}
\end{figure}

In \cref{fig:auger_ground}, we show the energy spectra of muons and neutrons together for an observation distance of more than 400\,m at $X_\text{det}=878\,$g/cm$^2$.
We avoid the region close to the shower core because in typical experiments it is dominated by signals from the electro-magnetic component.
We add as light-colored lines in \cref{fig:auger_ground} the spectra of muons and neutrons, with the same selections but in the presence of a ground layer of soil or other material at the observation level similar to the Pierre Auger site~\cite{PierreAuger:2015eyc}.
The difference between red and blue lines for the neutron spectra is the selection of time delays small enough to be recorded in typical cosmic-ray observatories, up to 21.5\,$\upmu$s.
As expected, the lowest neutron energies are removed with such a selection.
The minimal time delay, chosen as 1\,$\upmu$s, represents a typical time window to see delayed particles and has a smaller effect on the detectable neutron numbers: only the less abundant high energy part above about 10\,GeV is removed.
We can as well see that the modification of the energy spectra due to the ground only affects neutrons with a few MeV of kinetic energies.
However, as we will discuss here, this energy range is particularly interesting for detection and as such highlights the need to add and understand the ground level for detailed predictions.
Comparing the red with the gray spectra of \cref{fig:auger_ground}, it is clear that the number of neutrons (at these large distances and delays) is comparable to the number of muons.
These properties qualitatively agree with those reported by Linsley~\cite{Linsley:1984bnz} and others.

To gauge the possibility of detecting the neutron component, we perform calculations in two very simplified geometries, assessing the probability of neutron detection at given energies with plastic scintillators, and water/ice based Cherenkov detectors.
For these calculations, the \textsc{Fluka} fully-correlated pointwise neutron cross-section model, see~\cref{sec:fluka}, is exploited to be able to describe individual interactions in the detector materials.
In both cases the neutron angular distribution is assumed to be semi-isotropic since low-energy neutrons tend to be uniformly distributed in angle.

In the first case, an infinite, 1\,cm thick, plastic scintillator is exposed to neutrons and the resulting energy depositions scored whenever exceeding a threshold set at 100\,keV electron equivalent (the energy deposition of a minimum ionising particle (MIP) is at ${\sim}1.65$\,MeV electron equivalent).
The light response of the scintillator is assumed to follow the standard phenomenological Birk's law with an assumed quenching parameter of $10^{-2}$\,g/(MeV\,cm$^2$).
The computed detection probabilities should be taken as purely indicative, the actual efficiency will depend strongly on thresholds, the specific scintillator quenching parameters, and on the real-life setup, with reflection from surrounding materials and air likely to significantly increase the detection efficiency for low-energy neutrons.

In the second setup, an infinite, 120\,cm thick, water layer is irradiated with neutrons, and the corresponding energy deposition weighted for Cherenkov production efficiency, with a threshold set at two different levels,
$\sfrac{1}{300}$ and $\sfrac{1}{100}$ of the most probable signal for a vertically incident muon (VEM, Vertical Equivalent Muon) is taken.
For neutrons below the pion production threshold, the Cherenkov emission is mostly due to electrons from photons emitted in nuclear interactions, since no charged product can be above the Cherenkov production threshold.
The first threshold is below the Cherenkov signal produced by the 2.23\,MeV $\upgamma$ photons emitted in neutron capture on hydrogen, while the higher threshold is above this limit.
Therefore, with the higher threshold, disregarding the tiny fraction of neutrons absorbed on oxygen isotopes, there is an effective minimal neutron energy large enough to excite nuclear levels in oxygen.

Note that also for the water/ice Cherenkov detector the computed detection probabilities should be taken as purely indicative, the actual efficiency will depend strongly on the detailed assumptions, for example, the adopted detection threshold and the experimental setup, with reflection and showering from/into surrounding materials and air likely to increase the detection efficiency significantly.
The computed detection probabilities are listed in \cref{tab:neueff} for several incident neutron energies, for both the scintillator and water/ice-Cherenkov cases.

\begin{table}
\caption{Computed detection probabilities for semi-isotropic neutrons impinging on infinite slabs of 1\,cm thick plastic scintillator, and on 1.2\,m deep layer of water (see text for details about the setup).
The statistical errors are $\ll1\%$ of the computed probability for all probabilities above 0.1\%, ${<}10\%$ otherwise.}
\label{tab:neueff}
\begin{center}
\begin{tabular}{cccc}
\hline\hline
                   & \multicolumn{3}{c}{Detection probability (\%)}
\\
\cline{2-4}
Neutron            & \multicolumn{1}{c}{Scintillator}  
                   & \multicolumn{2}{c}{Water}
\\
Energy             & \multicolumn{1}{c}{Threshold}  
                   & \multicolumn{2}{c}{Threshold}
\\
(MeV)              & (100\,e-keV) & ($\sfrac{1}{300}$\,VEM) & ($\sfrac{1}{100}$\,VEM)
\\
\hline
0.0001             & $2.3{\times}10^{-2}$ & 13.7 & ${<}10^{-3}$
\\
0.001              & $1.0{\times}10^{-2}$ & 13.7 & ${<}10^{-3}$
\\
0.01               & $4.2{\times}10^{-3}$ & 13.7 & ${<}10^{-3}$
\\
0.1                & $1.3{\times}10^{-3}$ & 15.0 & ${<}10^{-3}$
\\
0.5                & ${<}10^{-3}$ & 18.5 & ${<}10^{-3}$
\\
0.7                & 4.65 & 20.1 & ${<}10^{-3}$
\\
1                  & 14.7 & 16.9  & ${<}10^{-3}$
\\
2                  & 17.1 & 25.1  & ${<}10^{-3}$
\\
3                  & 15.5 & 28.0  & ${<}10^{-3}$
\\
5                  & 12.4 & 29.0  & $4{\times}10^{-3}$
\\
10                 & 9.78 & 41.3  & 11.1
\\
20                 & 7.67 & 49.2  & 19.1
\\
30                 & 6.46 & 53.2  & 22.8
\\
50                 & 4.47 & 58.6  & 30.3
\\
100                & 2.87 & 61.8  & 37.5
\\
200                & 2.30 & 63.9  & 44.4
\\
500                & 2.31 & 75.3  & 52.3
\\
1000               & 2.55 & 83.2  & 79.7
\\
\hline\hline
\end{tabular}
\end{center}
\end{table}

Comparing the maximum of the (approximate) detection efficiency in scintillators at a few MeV with the energies of enhanced neutron fluxes for an added ground level in \cref{fig:auger_ground}, we can see that these energy ranges coincide.
While this increases the prospects of detecting neutrons in existing experiments without the need to extend the detection time, it also proves that the exact prediction of the neutron signals in a given experiment is challenging not only because of the shower physics but also due to the possible dependence of the detectable rate on external variable factors, like for example the ground humidity.

\section{Discussion}
\label{sec:discussion}

Neutrons produced in extensive air showers exhibit energy spectra and arrival-time distributions distinct from those of the electromagnetic and muonic shower components.
Main production channels are baryon pair production in hadronic interactions and disintegration of nuclei of air.
Both the hadronic interaction of shower particles with nuclei and the photo-disintegration of nuclei caused by the electromagnetic shower component are important sources of low-energy neutrons.
The energy loss of neutrons in the atmosphere is mainly driven by elastic and quasi-elastic interactions with target nuclei and leads to a characteristic $E^{-1}$ energy spectrum below kinetic energies of neutrons of about 10\,keV.
The length scale of the energy-loss processes, typically described by the so-called neutron-removal length, is about 100\,g/cm$^2$.

These distinct characteristics and the sheer abundance of neutrons make them secondary particles that are of potential interest to air-shower experiments.
The delay in the arrival times of the bulk of neutrons can serve as a very effective mean to identify their presence in time-resolved measurements of sufficient duration. 

We have also shown that an interplay between the energy in the hadronic component of air showers and attenuation results in an approximately linear scaling with primary energy of the number of potentially detectable neutrons arriving at observation depths typical of high-energy and ultra-high-energy cosmic-ray observatories.
This means that the number of neutrons increases faster with energy than that of muons.

A coincidence of similar production and attenuation effects also results in an unexpected but striking similarity of the number of potentially detectable neutrons for different hadronic shower primaries, as demonstrated for proton and iron showers.
The generally lower abundance of neutrons in photon-induced air showers may provide additional information to differentiate photons from hadronically-induced showers, if neutrons can be identified by their late arrival times.

To verify the predicted characteristics of the neutron cloud with measurements at a quantitative level, detailed detector simulations will be necessary to accurately account for the experimental setup and corresponding effects (e.g.\ quenching in scintillators, neutron reflection off of the ground).
Therefore, a general statement on the relevance of the neutron component in air-shower experiments is beyond the scope of this article.
Still, it can be noted that neutrons should be included in air-shower simulations to make sure the simulated ground signal includes also late particles.
Detector setups, with which only integrated charge information is produced, can be particularly sensitive to the details of late-arriving particles as discussed, for example, in \cite{Drescher:2005bg}.

A comparison of simulation predictions with existing neutron measurements, made with dedicated detectors taking data in coincidence with air-shower installations (see \cite{Stenkin:2023zau} for a recent review), is only possible if the detailed treatment of neutrons is implemented in the simulation chain.
Together with the challenges of neutron detection, this is probably the main reason why neutron data are typically not exploited in modern air-shower experiments.

\begin{acknowledgments}
It is our pleasure to acknowledge very fruitful and motivating discussions with our colleagues of the Pierre Auger Collaboration. 
We thank Alan A.\ Watson who has drawn our attention to the early work of Linsley~\cite{Linsley:1984bnz} and Hillas on this subject and provided us with an unpublished note by Hillas~\cite{Hillas:unpublished} on neutron production and its relevance for different air-shower detectors.
This research is supported in part by the Helmholtz International Research School for Astroparticle Physics and Enabling Technologies (HIRSAP), Helmholtz Association grant No.~HIRS-0009, and by the German Ministry for Education and Research, BMBF grant No.~05A2023VK4.
\end{acknowledgments}

\appendix*

\section{Kinematics of Elastic Scattering}
\label{app:elastic_scattering}

Let us recall the relations, which link the four-momentum transfer in a two-body elastic collision, $\tilde{q}$, to the energy transferred to a stationary target, and to the scattering angle in the center-of-mass system.
In the following $m_\text{n}$, $E_\text{kin,n}$, and $p_\text{lab,n}$ are the mass, momentum, and 
kinetic energy of the incoming projectile (neutron, n), respectively, $M_\text{t}$ the target mass, $T_\text{t}$ its recoil energy, $\sqrt{s}$, $p_\text{cms}$, and $\theta_\text{cms}$ the centre-of-mass energy, momentum, and scattering angle, respectively.
%
\begin{align}
T_\text{t} & = \frac{\tilde{q}^2}{2M_\text{t}},
\label{eq:ttoq2}
\\
p_\text{cms} & = p_\text{lab,n} \, \frac{M_\text{t}}{\sqrt{s}},
\\
s & = (M_\text{t} + m_\text{n})^2 + 2E_\text{kin,n} \, M_\text{t},
\\
\tilde{q}^2 & = \vec{q}^2_\text{cms} = 2 p_\text{cms}^2 (1 - \cos\theta_\text{cms}),
\label{eq:q2totheta}
\\
q^2_\text{max} & = 4 p_\text{cms}^2,
\\
T^\text{max}_\text{t} & = 2 \frac{p^2_\text{lab,n} \, M_\text{t}}{s}.
\end{align}
In the non relativistic regime ($E_\text{kin,n} \ll m_\text{n}$) we are interest in, the following approximate relations hold,
\begin{align}
E_\text{kin,n} & \approx \frac{p_\text{lab,n}^2}{2m_\text{n}},
\\
s & \approx (m_\text{n} + M_\text{t})^2,
\end{align}
and therefore
\begin{align}
T^\text{max}_\text{t} & = 4 E_\text{kin,n} \frac{m_\text{n}\,M_\text{t}}{(m_\text{n} + M_\text{t})^2},
\\
E^\text{min}_\text{kin,n} & \equiv
  E_\text{kin,n} - T^\text{max}_\text{t} =
  E_\text{kin,n} \left(\frac{M_\text{t} - m_\text{n}}{M_\text{t} + m_\text{n}}\right)^2 \equiv
  \alpha \, E_\text{kin,n}.
\end{align}

\section{Absolute Lateral Particle Densities}
\label{app:absolute_ldf}

\begin{figure}
\centering
\includegraphics[width=\columnwidth]{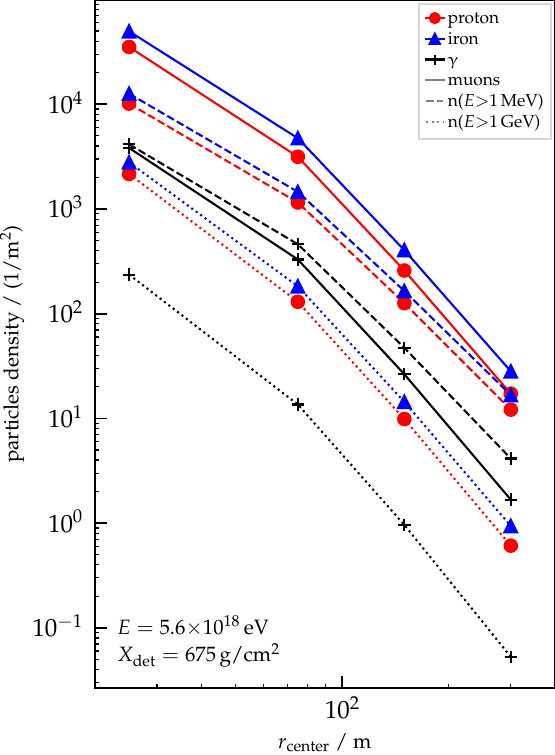}
\caption{Radial dependence of the muon and neutron distributions for different primaries at $E=5.6{\times}10^{18}$\,eV.
Proton showers are shown in red, iron in blue, and photon showers in black.
We integrate muons over the full energy range (solid), and show two energy selections for neutrons (dashed, dotted).
The binning is the same as presented in \cref{sec:simmethod} and plotted are the bin centers.}
\label{fig:ldf_like_he}
\end{figure}

\cref{fig:ldf_like_he} shows the absolute lateral particle densities as observed in our simulation setup for different primaries at $E=5.6{\times}10^{18}$\,eV.
As we have only access to the particle spectra in a limited set of radial intervals, we compute the lateral distributions shown here as integrals over time and energy for each of the radial intervals.
Due to the discrete energy binning used, these integrals, resulting in a number of particles above a minimal energy $N(E_\text{min})$, can be written as sum over the energy bins, as 
\begin{equation}
N(E_\text{min}) =
  \sum_i \sum_t \frac{\dd N_{t,i}}{\dd E} \, \Delta E_i \, \Theta(E_i - E_\text{min})
\end{equation}
with the minimal kinetic energy $E_\text{min}$ applied in the Heaviside function, and the observed spectrum in time and energy bin $N_{t,i}$.
To convert the energy spectrum into a particle count, the energy bin width $\Delta E_i$ is used.
To obtain the particle density $\rho$ we divide the particle number $N(E_\text{min})$ with the area of the segment covered by the radial bin with edges $r_1$ and $r_2$ as 
\begin{equation}
\rho = \frac{N(E_\text{min})}{\pi(r_2^2 - r_1^2)}.
\end{equation}
Due to the steeply falling lateral distribution of particles, the particular choices of radii $r_i$ for plotting in \cref{fig:ldf_like_he} are highly influential.
To avoid the explicit assumption of a lateral distribution, we take the arithmetic mean of the radii forming the bin, i.e.\ $r_\text{center}=(r_1 + r_2)/2$ and mark the values explicitly while guiding the eye with a line between the segments.

\bibliography{bibliography}

\end{document}